\documentclass[%
aip,jap,
 amsmath,amssymb,floatfix,
 preprint,%
]{revtex4-1}

\usepackage[utf8]{inputenc}
\usepackage[T1]{fontenc}
\usepackage{array}
\usepackage{amsmath,amssymb,amsfonts}
\usepackage{algorithmicx}
\usepackage{mathtools}
\usepackage{algorithm}
\usepackage{algpseudocode}
\usepackage{graphicx}
\usepackage{textcomp}
\usepackage{xcolor}
\usepackage{siunitx}
\usepackage{mathtools}
\DeclarePairedDelimiter\abs{\lvert}{\rvert}%

\def\BibTeX{{\rm B\kern-.05em{\sc i\kern-.025em b}\kern-.08em
    T\kern-.1667em\lower.7ex\hbox{E}\kern-.125emX}}
\begin{document}

\title{2-output Spin Wave Programmable Logic Gate
}

\author{Abdulqader Mahmoud}
\email{a.n.n.mahmoud@tudelft.nl}
\affiliation{Delft University of Technology, Department of Quantum and Computer Engineering, 2628 CD Delft, The Netherlands}

\author{Frederic Vanderveken}
\affiliation{KU Leuven, Department of Materials, SIEM, 3001 Leuven, Belgium}
\affiliation{Imec, 3001 Leuven, Belgium}

\author{Christoph Adelmann}
\affiliation{Imec, 3001 Leuven, Belgium}

\author{Florin Ciubotaru}
\affiliation{Imec, 3001 Leuven, Belgium}

\author{Sorin Cotofana}
\affiliation{Delft University of Technology, Department of Quantum and Computer Engineering, 2628 CD Delft, The Netherlands}

\author{Said Hamdioui}
\email{S.Hamdioui@tudelft.nl}
\affiliation{Delft University of Technology, Department of Quantum and Computer Engineering, 2628 CD Delft, The Netherlands}

\begin{abstract}
This paper presents a $2$-output Spin-Wave Programmable Logic Gate structure able to simultaneously  evaluate any pair of  AND, NAND, OR, NOR, XOR, and XNOR Boolean functions. Our proposal provides the means for fanout achievement within the Spin Wave computation domain and energy and area savings as two different functions can be simultaneously evaluated on the same input data. We validate our proposal by means of Object Oriented Micromagnetic Framework (OOMMF) simulations and demonstrate that by phase and magnetization threshold output sensing  \{AND, OR, NAND, NOR\} and \{XOR and XNOR\} functionalities can be achieved, respectively.  To get inside into the potential practical implications of our approach we use the proposed gate to implement a $3$-input Majority gate, which we evaluate and compare with state of the art equivalent implementations in terms of area, delay, and energy consumptions. Our estimations indicate that the proposed gate provides $33$\% and $16$\% energy and area reduction, respectively, when compared with spin-wave counterpart and $42$\% energy reduction while consuming $12$x less area when compared to a $15$ nm CMOS implementation. 
\end{abstract}

\maketitle

\section{Introduction}
During the past decades, the human society experienced an information technology revolution that resulted in a huge increase of easy available raw data, which processing requires efficient computing platforms ranging from high-performance clusters to simple Internet of Things (IoT) nodes \cite{data1,data2}. However, CMOS downscaling that provided the means to meet the data processing energy and performance requirements~\cite{ITRS} became more and more difficult due to various technological hurdles indicating  that Moore's Law will soon come to its end mainly because of: (i) leakage wall \cite{cmosscaling2,cmosscaling3}, (ii) reliability wall \cite{cmosscaling1}, and (iii) cost wall  \cite{cmosscaling1,cmosscaling2}. Therefore, to keep the pace with "exploding" market needs, novel alternative technologies are under investigation \cite{ITRS}. Among them Spin-Wave (SW) stands apart as one of the most promising avenue  \cite{survey1,survey2,ITRS} because it has: (i)  ultra low power consumption potential - SW based calculations are performed by means of SW interactions and do not require charge movements, 
(ii) acceptable delay, and (iii) high scalability - SW wavelengths can reach into the $nm$-range. Therefore, novel Spin Wave technology circuit design methodologies are of great interest.

Up to date different SW logic gates have been introduced, e.g.,  \cite{logic21,logic12,logic11,Magnonic_transistor,logic24,logic1,logic19,logic100,logic101}. The current controlled Mach-Zehnder interferometer was employed to construct a NOT gate, which is considered to be the first experimental work to implement logic gates using spin-waves \cite{logic21}. Afterwards, XNOR, NAND and NOR gates were implemented using Mach-Zehnder as well \cite{logic12,logic11,logic17}. Furthermore, a magnon transistor has been utilized to build an XOR gate by embedding two transistors in the Mach-Zehnder interferometer arms \cite{Magnonic_transistor}. Moreover, research was conducted to implement voltage controlled XNOR and NAND using two parallel re-configurable nano-channel magnonic devices \cite{logic24}.  Information encoding in SW phase rather than in SW amplitude also proposed \cite{logic1} and  buffer, inverter, AND, NAND, OR, NOR and XOR gate designs  introduced \cite{logic1}. Furthermore, a $3$-input majority gate design, which can perform $2$-input  AND and OR by assigning one of its inputs to $0$ or to $1$, respectively, was presented \cite{logic1} . In addition, OR and NOR were implemented using the cross structure \cite{logic19}. Furthermore, two experimental prototypes for majority gates were presented \cite{logic100} and \cite{logic101}. However, due to SW interaction way of operation, all of the reported logic gate designs cannot provide fan-out support, which is an essential gate feature for the effective implementation of larger circuits. Hence, if a SW gate output should serve as input for multiple following gates in the circuit it has to be replicated, leading not only to area overhead, but also to higher energy consumption.

This paper solves the above limitation and proposes a $2$-output Programmable Logic Gate (PLG). Depending on the design of the structure, the $2$ outputs can produce the same or different functions simultaneously. The main contributions of this work are:\\
\begin{itemize}
  \item Design of $2$-input $2$-output PLG: Two logic functions (including AND, NAND, OR, NOR, XOR and XNOR) can be implemented using a single $2$-output structure. For example, assuming that the gate inputs are $x$ and $y$ it can simultaneously provide both AND($x,y$) and XOR($x,y$).  
  \item Validation of gate functionality: Object Oriented Micromagnetic Framework (OOMMF) software is used to successfully validate the proposed gate behaviour for all considered  Boolean functions and input cases. 
  \item Demonstration of the superiority: The evaluations indicate that the proposed $2$-output gate saves $33$\% of energy and $16$\% of area without incurring any delay penalty when compared with functionally equivalent designs based on state-of-the-art spin-wave gates. Moreover, the proposed design outperforms $15$ nm CMOS in terms of energy and area by providing an energy reduction of $42$\% while consuming $12$x less area.
\end{itemize}

The rest of the paper is organized as follows. Section \ref{sec:Basics of spin-wave technology} provides basic spin-wave technology background. Section \ref{sec:Proposed programmable logic gate system design} describes the proposed Programmable Logic Gate design. Section \ref{sec:Simulation Setup and Experiments} introduces the OOMMF simulation setup and results while in Section \ref{sec:Discussion} we evaluate the proposed design, compare it with the state of the art, and discuss issues like scalability, variability, and thermal noise effects. Section \ref{sec:Conclusion} concludes the paper.

\section{SW technology background}
\label{sec:Basics of spin-wave technology}
A Spin Wave (SW) is the collective spin excitation in a magnetic system \cite{Magnonic_crystals_for_data_processing} and  this precessional motion of the magnetization is described by the Landau-Lifshitz-Gilbert equation \cite{LL_eq}\cite{G_eq} as:

\begin{equation} \label{eq:1}
\frac{d\vec{m}}{dt} =-\abs{\gamma} \mu_0 \left (\vec{m} \times \vec{H}_{eff} \right ) + \frac{\alpha}{M_s} \left (\vec{m} \times \frac{d\vec{m}}{dt}\right ),
\end{equation}
where $\alpha$ is the damping factor, $\gamma$ the gyromagnetic ratio, $M_s$ the saturation magnetization, $\vec{m}$ the magnetization and $H_{eff}$ the effective magnetic field given by
\begin{equation} \label{eq:2}
H_{eff}=H_{ext}+H_{ex}+H_{demag}+H_{ani}+H_{ani}^{sh},
\end{equation}
where $H_{ext}$ is the external field, $H_{ex}$ the exchange field, $H_{demag}$ the demagnetizing field, $H_{ani}$ the magneto-crystalline field and $H_{ani}^{sh}$ the shape field.

\begin{figure}[t]
\centering
  \includegraphics[width=\linewidth]{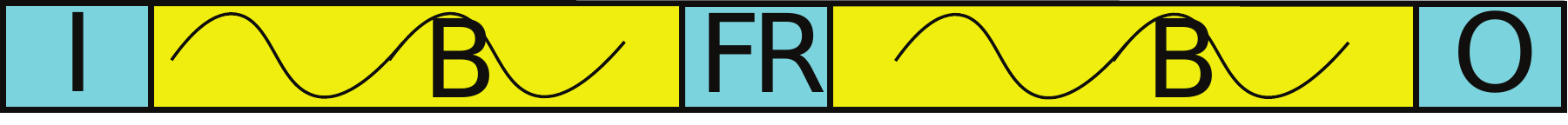}
  \caption{Generic Spin Wave Device}
  \label{fig:fig1}
\end{figure}

\begin{figure}[b]
\centering
  \includegraphics[width=0.7\linewidth]{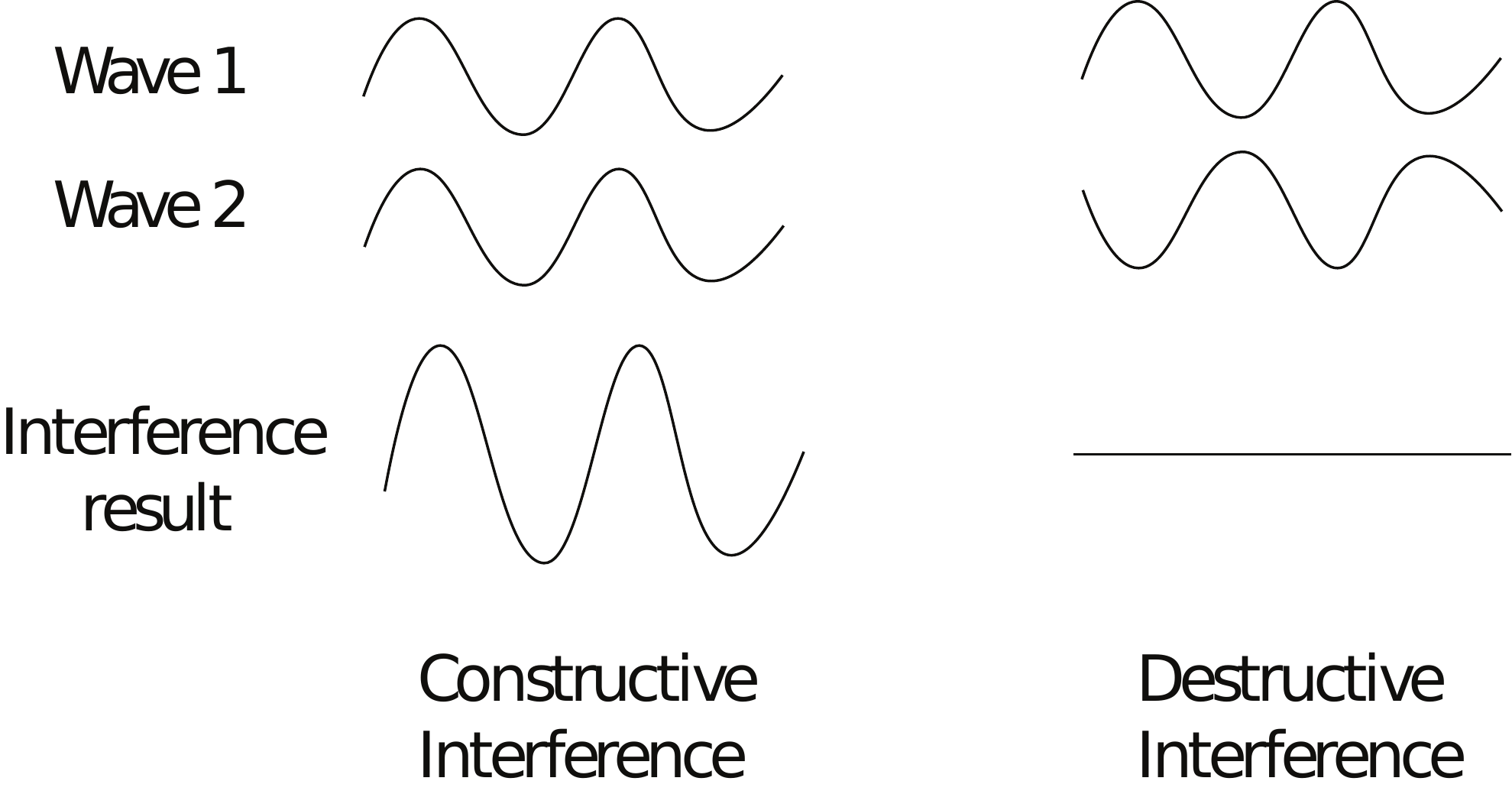}
  \caption{SW Constructive and Destructive Interference}
  \label{fig:fig2}
\end{figure}

The main interactions which give rise to spin waves are the exchange and dipolar interaction for respectively short and long wavelength spin waves \cite{Magnonic_crystals_for_data_processing}. Depending on SW propagation direction relative to the orientation of the magnetization and effective magnetic field, different SW types can be excited and each has its own characterstics. Three main spin-wave regimes exist depending on the wavelength: exchange spin-waves, exchange-dipole spin-waves, and dipole (magnetostatic) spin-waves. There is a single dispersion relation, which represents the SW frequency as a function of the wavevector $k$ \cite{dispersionrelation}. While all three regimes are captured by a single relation this is not a linear function as the curvature can change depending on the dipolar/exchange contribution. Additionally, depending on the relative orientation between the wave propagation direction and the magnetization different spin wave types exist, namely Magnetostatic Surface spin-wave (MSSW), Backward Volume Magnetostatic spin-wave (BVMSW), and Forward Volume Magnetostatic spin-wave (FVMSW) each one characterised with its own dispersion relation \cite{Magnetostatics_ref3}. We note that FVMSW in-plane propagation is isotropic and as a result the same wave number is excited in all directions, which is not the case for other types. In view of this FVMSWs are the most promising from the circuit design prospective \cite{Magnetostatics_ref3} and are utilized  in this paper.


As depicted in Figure \ref{fig:fig1}, a spin wave device consists of $4$ regions: excitation region I (where SWs are excited), waveguide B (where SW propagate), functional region FR (where SWs can be manipulated), and detection stage O (where tha output SW is detected). During the excitation stage, information can be encoded into the amplitude ($A$) and phase ($\phi$) of  spin-waves at different frequencies ($f$) \cite{parallel_data_processing1,counter}. This makes it possible to use spin-waves as data carriers and their interaction as data processing mechanism to enable parallelism in SW circuits. If multiple waves co-exist in the same waveguide, their interactions is based on the interference and superposition principles. When SWs interfere in the waveguide, their interaction can be constructive or destructive depending on their phase difference. Figure \ref{fig:fig2} illustrates these two cases. Two spin waves having the same wavelength ($\lambda$) constructively interference if they have the same phase ($\Delta \phi=0$) and destructively  if they are out of phase ($\Delta \phi=\pi$). If more than two waves coexist in the waveguide, then their interference results is based on a majority decision. By assuming that input SWs can have phases $\phi=0$ or $\phi=\pi$ their interference results in a SW with $\phi=0$ if the majority of inputs have $\phi=0$, and in a SW with $\phi=\pi$ otherwise. This implies that SW interaction provides natural support for, e.g., single gate $3$-input majority function implementation, which in a CMOS technology based Boolean logic based implementation requires $18$ transistors \cite{logic1,logic9}.

\section{$2$-input $2$-output Programmable Logic Gate}
\label{sec:Proposed programmable logic gate system design}

This section describes the proposed gate structure and discusses different logic gate embodiments.

\subsection{Proposed Programmable Logic Gate Structure}
\label{subsec:Proposed system structure}
\begin{figure}[t]
\centering
  \includegraphics[width=0.3\linewidth]{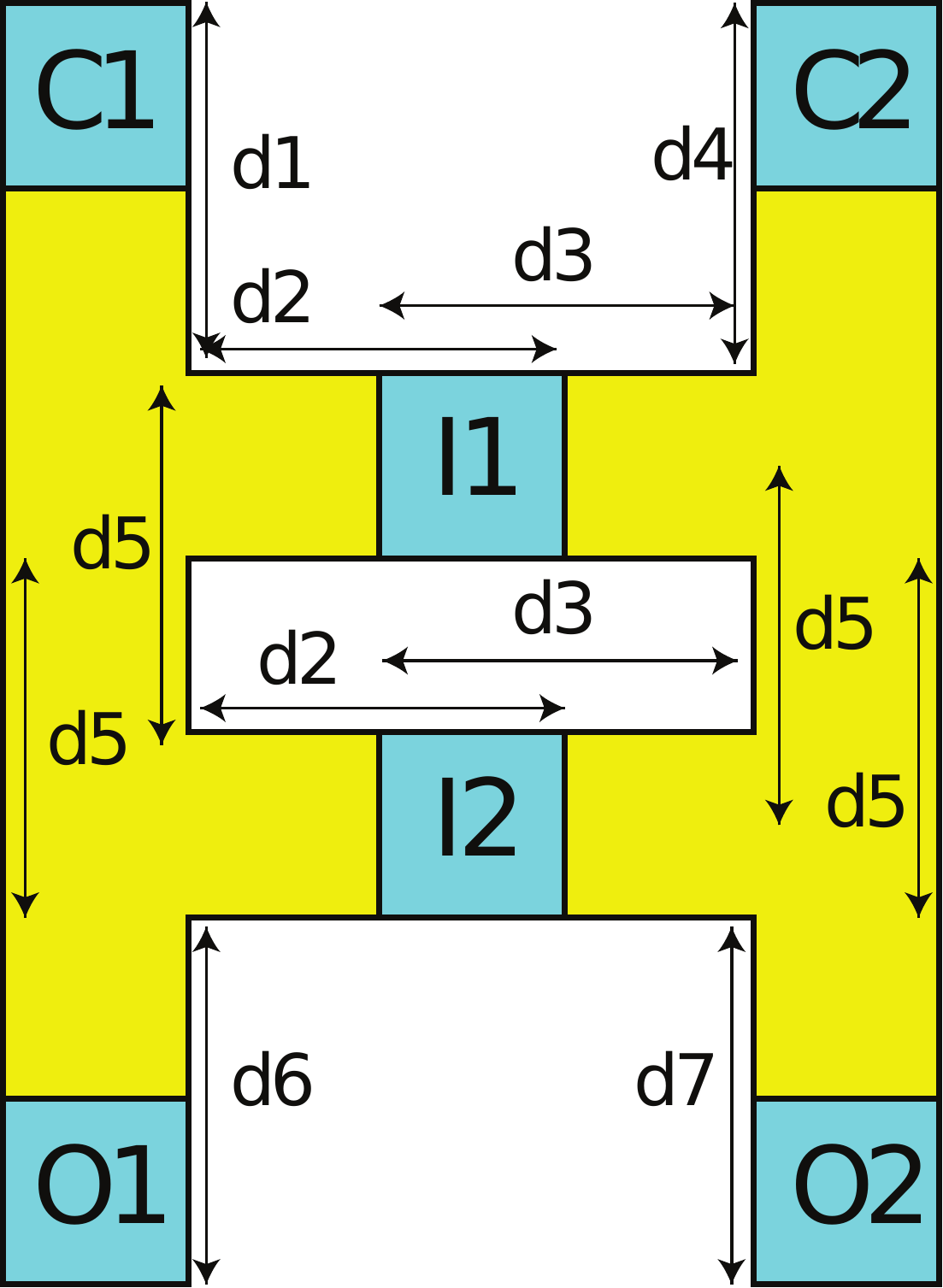}
  \caption{2-input 2-output SW Programable Logic Gate}
  \label{fig:structure2}
\end{figure}

Figure \ref{fig:structure2} graphically depicts the topology of the proposed programmable logic gate that has a ladder shape structure. It has 2 inputs ($I_1$ and $I_2$) placed on the stair's stpdf and two control signals $C_1$ and $C_2$ located at the ladder top that control the way the input spin-waves are interacting. Further, the outputs $O_1$ and $O_2$ are obtained at the end of both columns. Both input SWs generation and output SWs detection are performed by means of magnetoelectric (ME) cells or other transducers \cite{Magnonic_crystals_for_data_processing} able to transform voltage (current) encoded logic values into SWs and the other way around. 
To obtain the correct expected output results the distances between SW interference points must be accurately determined as multiples of the SW wavelength $\lambda$. In particular, if $d_i=n \lambda$ (where $i \in \{1,2,3,4,5\}$, and $n=0,1,2, \ldots$), the SWs are constructively interfering if they have the same phase ($\Delta \phi=0$) and destructively  if they are out of phase ($\Delta \phi=\pi$). On the other hand, when $d_i=(n+\frac{1}{2}) \lambda$, SWs are destructively interfering if $\Delta \phi=0$ and constructively if $\Delta \phi=\pi$.  
To correctly read the outputs, the distance $d_6$ and $d_7$ must be multiples of  $\lambda$.  The multiplicity factor  determines whether the direct or inverted output value is made available. Reading a non-inverted output at $O_1$ and $O_2$ requires $d_6 = d_7 =  n \lambda$, while if the output complement is of interest, $O_1$ and $O_2$ should be positioned at  $d_6 = d_7 = (n+\frac{1}{2}) \lambda$. 
In addition, there are two ways to detect the output: (i) Detection based on the phase, and (ii) Detection based on the threshold \cite{logic1}. Phase detection method is based on a predefined phase reading, i.e., if the detected spin wave has a phase of $\pi$, then the result is logic $1$, and a logic $0$ if the detected spin wave has a phase of $0$. However, threshold detection method is based on a predefined threshold, i.e., if the output magnetization is greater than or equal to a certain threshold, then the gate output is $0$ and $1$ otherwise.

\subsection{Logic Function Programming}
\label{subsec:Different logic gates implementation}

The logic function performed by the structure in Figure \ref{fig:structure2} depends on the detection mechanism as follows:  (i) If  phase detection is utilized the gate can evaluate (N)AND and (N)OR operators and (ii) if  threshold detection is in place the structure implements X(N)OR. 


To explain the programmable $2$-input gate operation based on phase detection, we consider the structure in Figure \ref{fig:structure2} under the assumption that  the control signals are $C_1=0$ and $C_2=1$. If a logic $0$ is applied on both $I_1$ and $I_2$, spin-waves with the same frequency, wavelength, and phase are excited in both horizontal waveguides. In the first stage, SW generated at $I_1$ constructively interferes with the one generated at $C_1$. The resulting SW continues the propagation downwards in the waveguide and eventually constructively interferes with the spin wave emitted by $I_2$, which results in a logic $0$ at the output cell $O_1$ because the interference results in a SW with phase of $0$. If $I_1=0$ and $I_2=1$, the waves (from $C_1$ and $I_1$) interfere constructively and the resulted SW propagates downwards to destructively interfere with the wave from $I_2$, which results in a logic $0$ at the output cell $O_1$ because the interference results in a SW with phase of $0$. When $I_1=1$ and $I_2=0$, the waves from the $I_1$ and $C_1$ interfere destructively, which diminishes the resulted SW energy to a minimum if not make it vanish completely. Thus, the wave emitted by $I_2$ is the only one  still present in the waveguide and propagates to the output resulting in a logic $0$ at $O_1$ because the interference results in a SW with phase of $0$. Finally, when $I_1=1$ and $I_2=1$, the same interference as in the previous case happens but since $I_2$ is now logic $1$, the output is logic $1$ at $O_1$ because the interference results in a SW with phase of $\pi$. Thus, the left side of the structure behaves as an AND gate. Following the same line of reasoning, one can easily demonstrate that the right side behaves as an OR gate. By changing the control signals values to $C_1=0$ and $C_2=0$, $C_1=1$ and $C_2=0$, and $C_1=1$ and $C_2=1$ the gate functionality is changed to (AND, AND), (OR,AND), and (OR,OR), respectively. Thus the gate can provide fanout of $2$ OR/AND behaviour but also the parallel evaluation of both AND and OR over the same input values. 
Furthermore, the gate can also produce NOR and NAND if the outputs are read at distances $d_6$ and $d_7$ equal to $(n+\frac{1}{2}) \lambda$. Thus the gate is capable to evaluate a rich set of function combinations, e.g., (NAND, NAND), (NOR,NAND), (NOR,NOR), (NAND, AND), (NOR,AND), (NOR,OR).

If thresholding-based output detection is utilized, the performed functions change into XOR and XNOR. This relates to the fact that in this case the output SW phase is ignored and the output logic value is asserted based on the SW energy level or Magnetization Spinning Angle (MSA) which can be calculated as: 
\begin{equation}
MSA = \arctan \left(\frac{\sqrt{(\overline{m_x})^2+(\overline{m_y})^2}}{M_s}\right),
\label{equ:MSA}
\end{equation} 
where $\overline{m_x}$ and $\overline{m_y}$ are the $x$ and $y$  magnetization components, respectively. 

We also note that, as output detection methods are not mutually exclusive, one output can be read in phase and the other one by thresholding, which enables the parallel evaluation of mixed function pairs, e.g., (AND,XOR), (OR,XOR).

\begin{table}[t]
\caption{Parameters}
\label{table:1}
\centering
  \begin{tabular}{|c|c|}
    \hline
    Parameters & Values \\
    \hline
    Magnetic saturation $M_s$ & $1.1$ $\times$ $10^6$ A/m \\
    \hline
    Perpendicular anisotropy constant $k_{ani}$ & $8.3177$ $\times$ $10^5$ J/$m^3$\\
    \hline
    damping constant $\alpha$ & $0.004$ \\
    \hline
    Waveguide thickness $t$ & $1$ nm \\
    \hline
    Exchange stiffness $A_{exch}$ & $18.5$ pJ/m \\
    \hline
  \end{tabular}
\end{table}

\section{Simulation Setup and Experiments}
\label{sec:Simulation Setup and Experiments}
To validate the behaviour of the proposed SW programable gate we make use of an Object Oriented MicroMagnetic Framework (OOMMF) \cite{OOMMF} based simulation platform.

During the OOMMF simulations we made use of the parameters  summarized in Table \ref{table:1} \cite{parameters} and to automate the simulation process we developed a Tckl/Tk script. To demonstrate the functionality of the proposed structure, we considered $Fe_{60}Co_{20}B_{20}$ magnetic waveguides of $50$ nm width, with a perpendicular magnetic anisotropy field greater than the magnetic saturation, thus no external magnetic field is required \cite{parameters}. (N)AND/(N)OR and X(N)OR gates for waveguide width $w = 50$ nm are instantiated based on the proposed structure. To determine the SW frequency, we have chosen a SW wavelength $\lambda = 110$ nm, which means that $d_1 = d_2 = d_3 = d_4 = d_5 = d_6 = d_7= 110$ nm, and from the dispersion relation calculated based on the parameters in Table \ref{table:1} and wavelength, the SW frequency is determined as $f = 9$ GHz at a spin wave number $k=2\pi/\lambda=57$ rad/$\mu$m.

\begin{figure}[t]
\centering
  \includegraphics[width=\linewidth]{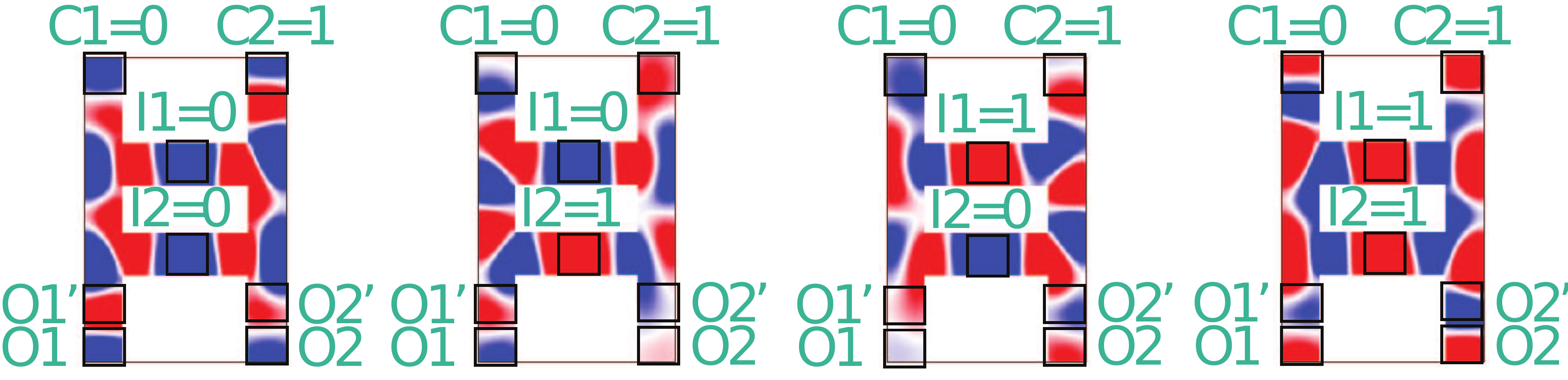}
  \caption{Fan-in of $2$ AND/OR Gate OOMMF Simulation}
  \label{fig:results1}
\end{figure} 
\subsection{Phase Detection based AND/OR gate} 
Figure \ref{fig:results1} presents the simulation results for a $2$-input gate with $C_1=0$ and $C_2=1$. The color code in the figure is: Red (dark) represents logic $1$ and Blue (light) represents logic $0$. 
The visual inspection of Figure \ref{fig:results1} reveals that the structure simultaneously produces AND and OR over the two inputs. $O_1$ provides the AND function, i.e., $O_1=0$ for the input combinations ($I_1I_2=00$,  $I_1I_2=01$,  $I_1I_2=10$), and  $O_1=1$ for $I_1I_2=11$. Similarly, $O_2$ produces the OR function, i.e., $O_2=1$ for the input combinations ($I_1I_2=01$,  $I_1I_2=10$,  $I_1I_2=11$), and  $O_1=0$ for $I_1I_2=00$. Note that $O_1$ and $O_2$ are placed at $d_6=d_7=110$nm ($n=1$) such that direct functions are obtained. The Figure also indicate that NAND and NOR functions are obtained at outputs $O_1'$ and $O_2'$ by just shifting the output reading points by $\lambda$/2 such that $d_6=d_7=55$nm ($n=0$).
Hence, Figure \ref{fig:results1} demonstrates the correct behaviour of the proposed $2$-input $2$-output logic gates. However, as one can observe in the Figure input value depended output signal strengths are obtained. For example, $O_2$ is weaker than $O_1$  when the inputs are $(0,0)$ because in the left column two constructive interferences take place (resulting in a strong output), while in the right column one destructive and one constructive interference occur (resulting in a weaker output). Thus, the output detection mechanism must be able to deal with such variability phenomena.

\subsection{Threshold Detection based XOR/XNOR gate} 

Table \ref{table:3} presents the normalized Magnetization Spinning Angles (MSA) values at $O_1$ and $O_2$ for $C_1=C_2=0$ and all possible $I_1$ and $I_2$ input values. The MSA values in the Table are computed based on Equation (\ref{equ:MSA}) and normalized  with respect to the highest MSA,  which in this case is obtained when $I_1 I_2=00$. Note that the results for the other possible control input combinations, i.e., $C_1 C_2$=$01$, $C_1 C_2$=$10$, and $C_1 C_2$=$11$, are similar to those obtained for $C_1 C_2=00$. 

The basic idea behind the threshold based output value interpretation is to define an appropriate MSA value $MSA_{th}$ and, e.g., classify the gate output as $0$ if its MSA value is larger than $MSA_{th}$ and $1$ otherwise.  
By applying this principle on the Table \ref{table:3} values and choosing $MSA_{th}=0.41$ the gate outputs will be both $0$ if $I_1 I_2=00$ and $I_1 I_2=11$ and $1$ otherwise, which means that the gate provides the XOR functionality.  If the detection rule is changed such that logic $1$ is reported when the normalized MSA value is larger than $MSA_{th}$ and logic $0$ otherwise, the proposed structure evaluates an XNOR function. Thus, in this case the output reading location is not relevant as the inverted version of the output is obtained by switching the thresholding rule.


\begin{table}[t]
\caption{$2$-input $2$-output Gate Normalized Output MSAs}
\label{table:3}
\centering
  \begin{tabular}{|c|c|c|c|c|}
    \hline
   \multicolumn{3}{|c|}{Cases} & $O_1/I_1$  & $O_2/I_1$ \\ \hline
    $C_1=C_2$ & $I_2$ & $I_1$& & \\
    \hline
    $0$ & $0$ & $0$ & $1$ & $1$ \\
    \hline
    $0$ & $0$ & $1$ & $0.28$ & $0.28$ \\
    \hline
    $0$ & $1$ & $0$ & $0.37$ & $0.37$ \\
    \hline
    $0$ & $1$ & $1$ & $0.45$ & $0.45$ \\
    \hline
  \end{tabular}
\end{table}

\begin{figure}[b]
\centering
  \includegraphics[width=\linewidth]{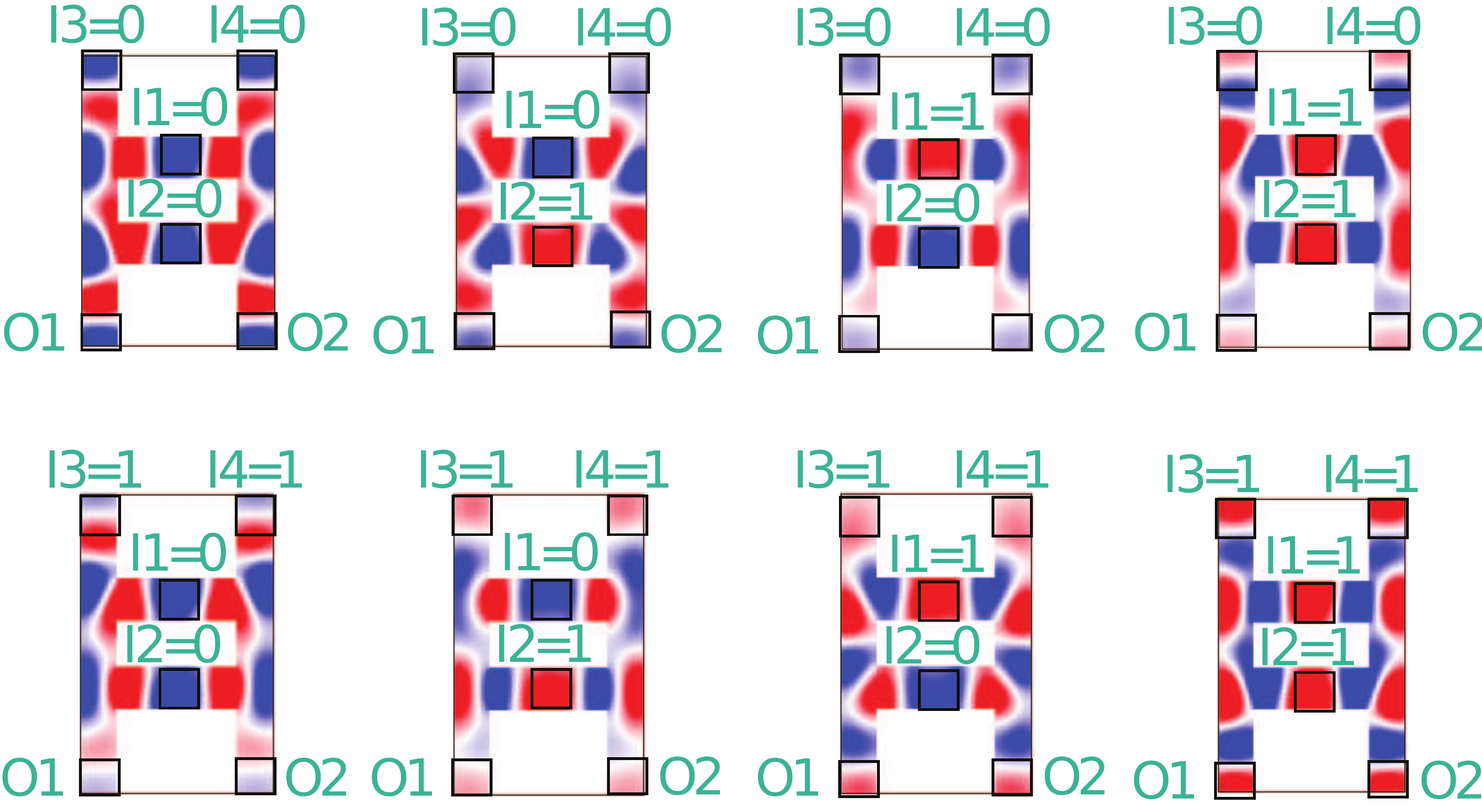}
  \caption{3-input Majority Gate}
  \label{fig:MAJ}
\end{figure} 

\section{Performance Evaluation and Discussion}
\label{sec:Discussion}
To asses the implications of our proposal, we make use of the proposed gate to implement a fanout of $2$ $3$-input majority (MAJ3) gate, evaluate its area, delay, and energy consumption and compared it with  the state of the art SW  \cite{Excitation_table_ref16}, and $15$ nm CMOS based counterparts. 

To this end we instantiated a MAJ3  gate design with a waveguide width of $48$ nm and $\lambda = 96$~nm and validated it by means of OOMMF simulations as presented in Figure \ref{fig:MAJ}. For a fair comparison with the MAJ3 implementation in \cite{Excitation_table_ref16}, we made use of the same assumptions: (i) ME cells are utilized for SW excitation and detection,  (ii) ME cell parameters are: Area = $48$~nm $\times$ $48$~nm area, Energy=$i \times C_{ME} \times V_{ME}^2$, where $i$ is the number of excitation cells, $C_{ME}=1$ fF, $V_{ME}=119$ mV), and Switching Delay = $0.42$~ns , (iii) We moved the output locations of our design and placed them immediately after the interference points, i.e., $d_6=d_7=0$, (iv) MAJ3 gate outputs are directly driving the following SW gates, thus no delay and energy overhead is accounted for the ME cells at the gate output, and (v) The spin-wave through the waveguide propagation delay  is negligible. Note that due to the SW technology early stage of development some of these assumptions might not accurately reflect the physical reality, but their discussion is out of the scope of this paper. 

In addition, in order to compare with CMOS, we evaluated a $3$-input Majority gate implemented with two NAND gates and one OR-AND-Invert (OAI) gate in $15$ nm technology at $V_{dd} = 0.8$ V, $25^{\circ}C$, and an output load capacitance of $20$ fF.

Table \ref{table:6} presents our evaluation results, it indicates that while the proposed SW gate is $14$x slower than the CMOS counterpart it provides a $42$\% energy consumption reduction while requiring $12$x  smaller area. The Table also indicate that the Majority gate in \cite{Excitation_table_ref16} is slightly more energy efficient. However, the design in \cite{Excitation_table_ref16} can only provide a single output, which means that if multi-output is desired replication is required which result in area and energy overhead. For example, if $2$ outputs are required when using the design in \cite{Excitation_table_ref16}, the structure must be replicated twice, which is doubling the energy consumption and area to  $86.6$ aJ and $0.0691$ $\mu m^2$, respectively. Given that our design consumes $56.6$ aJ, and requires $0.0576$ $\mu m^2$, it provides  $33$\%, $16$\% energy and area reduction, respectively, without any delay overhead. \\

\begin{table}[t]
\caption{Performance Comparison}
\label{table:6}
\centering
  \begin{tabular}{|>{\centering}m{9em}|>{\centering}m{6em}|>{\centering}m{5em}|>{\centering}m{5em}|}
    \hline
     & CMOS  &  SW \cite{Excitation_table_ref16} & SW \tabularnewline \hline
     Technology &  15 nm CMOS &  SW & SW \tabularnewline
    \hline
     Implemented function &  $2$-output MAJ3  & $1$-output MAJ3  & $2$-output MAJ3 \tabularnewline
    \hline
     Number of used cell & $16$ transistors &  $4$ ME cells & $6$ ME cells \tabularnewline
    \hline
     Fanout capability & \textgreater $2$ &  $1$ & $2$ \tabularnewline
    \hline
     Delay (ns) &  $0.031$ &  $0.42$ & $0.42$ \tabularnewline
    \hline
     Energy (aJ) &  $98$ &  $43.3$ & $56.6$ \tabularnewline
    \hline
     Area ($\mu m^2$) &  $0.688$ &  $0.0346$ & $0.0576$ \tabularnewline
    \hline
  \end{tabular}
\end{table}

In the remainder of this section we briefly discuss issues related to the scalability of our proposal and some practical implementation aspects.\\

\begin{figure}[t]
\centering
  \includegraphics[width=0.3\linewidth]{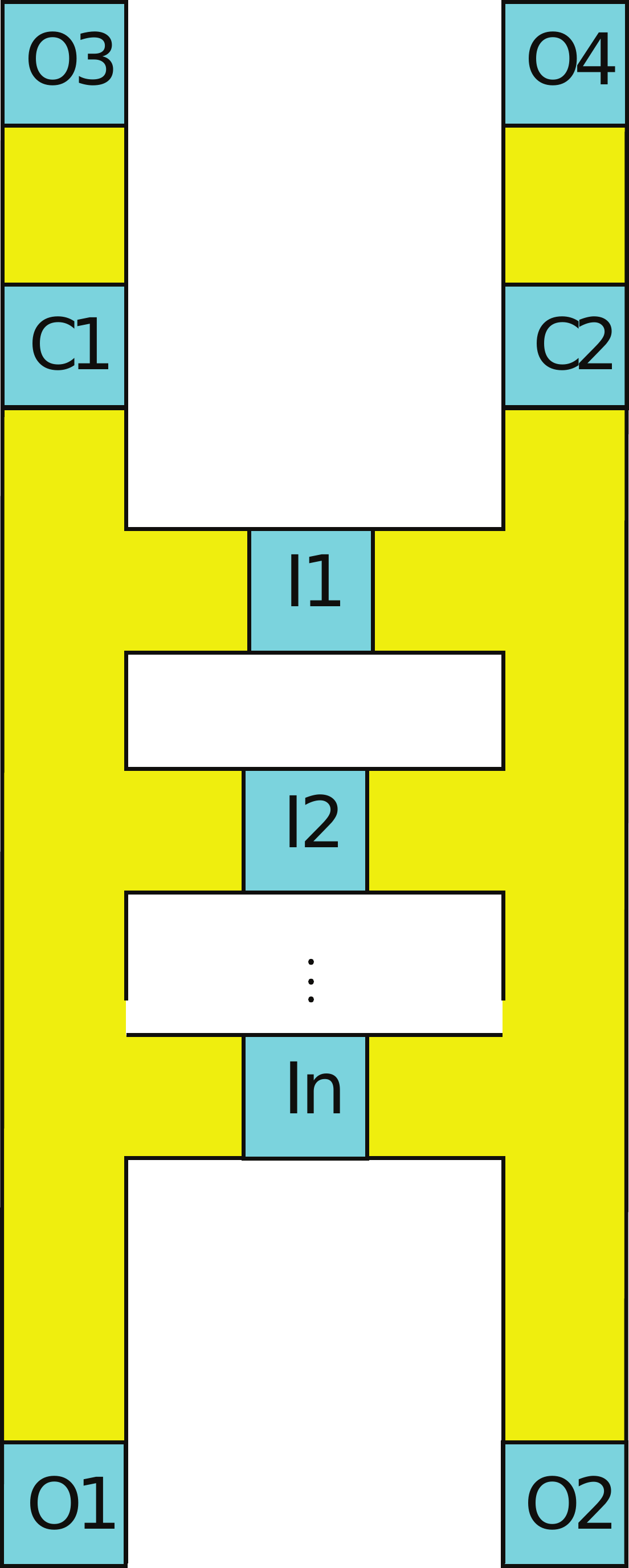}
  \caption{Multi-input Multi-output Programmable Logic Gate}
  \label{fig:FIFO}
\end{figure}

\noindent\textbf{Logic Scalability:}
The proposed structure is certainly scalable in terms of inputs and outputs. Additional inputs can be added by increasing the number of ladder stpdf. However, as the number of inputs increases, the inputs nearer to the outputs must be excited at lower energy to compensate for the potential degradation due to the damping effect in the waveguides of the SW inputs traveling from more faraway inputs. If the SW propagation delay is neglected the gate delay is number of inputs independent, while the energy increases linearly as the it is proportional to the number of inputs. In addition, the area of the proposed structure increases linearly with the number of inputs as the structure length increases by $w+\lambda$ when an input is added to the structure and the structure width is constant. In term of outputs, the scalability is limited by the number of columns. However, because SW propagates in all directions the two columns can be vertically extended such that outputs (up to $4$) can be read both at the top and bottom of each column. Thus the number of inputs and outputs of the proposed structure can be increased as depicted in Figure \ref{fig:FIFO}, but the detailed design of such a structure is out of the scope of this paper. \\

\noindent\textbf{Geometrical Scaling:} To examine the effect of the waveguide width scaling on the functionality of the proposed structures, two additional designs with waveguide widths of $30$ nm and $75$ nm were instantiated and validated by means of OOMMF simulations. We noticed that the width scaling does not detrimentally affects the functionality of the proposed structures and that when the waveguide width increases, the output MSA values increase as larger ME cells can excite stronger spin-waves. \\

\noindent\textbf{Balanced Spin Wave Strength:} It was noticed that the controls and data inputs contribute differently to the outputs. This is related to the fact that the control inputs have a larger contribution to the outputs than the data inputs as they have a direct straight path to the output, while the data inputs have bent region at the edges. Therefore, $C_1$ and $C_2$ must be excited at lower energy than $I_1$ and $I_2$ to balance the spin waves strength, and guarantee that the gate functions correctly. Furthermore, it was observed that all inputs $C_1$, $C_2$, $I_1$ and $I_2$ affect both outputs $O_1$ and $O_2$. Thus, $C_1$ has an effect on $O_2$, and $C_2$ has an effect on $O_1$, which might create a problem when different gates are captured at both outputs $O_1$ and $O_2$. Therefore, to guarantee a proper gate functioning correctly, it must be ensured that $C_1$, $I_1$ and $I_2$ contribute more on output $O_1$ compared to $C_2$, and that the contribution of $C_2$, $I_1$ and $I_2$ on output $O_2$ is more than the contribution of $C_1$. \\

\noindent\textbf{Variability:} The main purpose of this paper has been to propose the concept and validate it by means of micromagnetic simulations under ideal conditions even if it is obvious that waveguide dimension variations, edge roughness, and spin wave strength variation might affect the gate functionality.  However, given the actual development of the SW technology such aspects cannot be investigated for the time being but we certainly consider them as future work when relevant technology data became available. \\

\noindent\textbf{Thermal Noise:} Generally speaking the thermal noise is a  circuit design crucial issue. However, for spin wave technology, the Curie temperature, which is the temperature at which the magnetic properties of the material are changing, is high for the ferromagnetic materials, e.g., Curie temperature for CoFeB=1000 K \cite{Magnonic_crystals_for_data_processing}). So, it is expected that limited temperature variations over the room temperature do not fundamentally affect the gate behaviour. However, further investigations on the thermal impact are part of planned future work. 

\section{Conclusions}
\label{sec:Conclusion}
We introduced and evaluated by means of Object Oriented Micromagnetic Framework (OOMMF) simulations a novel $2$-input $2$-output Spin Wave based Programmable Logic Gate with a ladder shape structure capable to evaluate any pair of  AND, NAND, OR, NOR, XOR, and XNOR Boolean functions.  To get inside into the potential practical implications of our proposal we made use of such a gate to instantiate a $3$-input Majority gate, which we evaluated and compared with state of the art equivalent implementations in terms of area, delay, and energy consumptions. Our estimates indicated that the proposed gate provides $33$\% and $16$\% energy and area reduction, respectively, when compared with spin-wave counterpart and $42$\% energy reduction while consuming $12$x less area when compared to a $15$ nm CMOS implementation.

\section*{Acknowledgement}
This work has received funding from the European Union's Horizon 2020 research and innovation program within the FET-OPEN project CHIRON under grant agreement No. 801055.


\begin{thebibliography}{0}%
\makeatletter
\providecommand \@ifxundefined [1]{%
 \@ifx{#1\undefined}
}%
\providecommand \@ifnum [1]{%
 \ifnum #1\expandafter \@firstoftwo
 \else \expandafter \@secondoftwo
 \fi
}%
\providecommand \@ifx [1]{%
 \ifx #1\expandafter \@firstoftwo
 \else \expandafter \@secondoftwo
 \fi
}%
\providecommand \natexlab [1]{#1}%
\providecommand \enquote  [1]{``#1''}%
\providecommand \bibnamefont  [1]{#1}%
\providecommand \bibfnamefont [1]{#1}%
\providecommand \citenamefont [1]{#1}%
\providecommand \href@noop [0]{\@secondoftwo}%
\providecommand \href [0]{\begingroup \@sanitize@url \@href}%
\providecommand \@href[1]{\@@startlink{#1}\@@href}%
\providecommand \@@href[1]{\endgroup#1\@@endlink}%
\providecommand \@sanitize@url [0]{\catcode `\\12\catcode `\$12\catcode
  `\&12\catcode `\#12\catcode `\^12\catcode `\_12\catcode `\%12\relax}%
\providecommand \@@startlink[1]{}%
\providecommand \@@endlink[0]{}%
\providecommand \url  [0]{\begingroup\@sanitize@url \@url }%
\providecommand \@url [1]{\endgroup\@href {#1}{\urlprefix }}%
\providecommand \urlprefix  [0]{URL }%
\providecommand \Eprint [0]{\href }%
\providecommand \doibase [0]{http://dx.doi.org/}%
\providecommand \selectlanguage [0]{\@gobble}%
\providecommand \bibinfo  [0]{\@secondoftwo}%
\providecommand \bibfield  [0]{\@secondoftwo}%
\providecommand \translation [1]{[#1]}%
\providecommand \BibitemOpen [0]{}%
\providecommand \bibitemStop [0]{}%
\providecommand \bibitemNoStop [0]{.\EOS\space}%
\providecommand \EOS [0]{\spacefactor3000\relax}%
\providecommand \BibitemShut  [1]{\csname bibitem#1\endcsname}%
\let\auto@bib@innerbib\@empty
\end{thebibliography}%


\begin{thebibliography}{00}
\bibitem{data1} N. D. Shah, E. W. Steyerberg, and D. M. Kent, “Big Data and Predictive Analytics Recalibrating Expectations,” JAMA, 2018.
\bibitem{data2} R. L. Villars, C. W. Olofson, and M. Eastwood, “Big data What it is and why you should care,” IDC, 2011.
\bibitem{ITRS} S. Agarwal, G. Burr, A. Chen, S. Das, E. Debenedictis, M. P. Frank, P. Franzon, S. Holmes, M. Marinella, and T. Rakshit, “International Roadmap of Devices and Systems 2017 Edition: Beyond CMOS Chapter.” Sandia National Lab.(SNL-NM), Albuquerque, NM (United States), Tech. Rep., 2018.
\bibitem{cmosscaling2} D. Mamaluy and X. Gao, “The Fundamental Downscaling Limit of Field Effect Transistors,” Applied Physics Letters, vol. 106, no. 19, p. 193503, 2015.
\bibitem{cmosscaling3} B. Hoefflinger, Chips 2020: A Guide to the Future of Nanoelectronics. Springer Science and Business Media, 2012.
\bibitem{cmosscaling1} N. Z. Haron and S. Hamdioui, “Why is CMOS Scaling Coming to an End?” in Design and Test Workshop, 2008. IDT 2008. 3rd International. IEEE, 2008, pp. 98–103.
\bibitem{survey1} K. Bernstein, R. K. Cavin, W. Porod, A. Seabaugh, and J. Welser, “Device and Architecture Outlook for Beyond CMOS Switches,” Proceedings of the IEEE, vol. 98, no. 12, pp. 2169–2184, Dec 2010.
\bibitem{survey2} D. E. Nikonov and I. A. Young, “Overview of Beyond-CMOS Devices and a Uniform Methodology for their Benchmarking,” Proceedings of the IEEE, vol. 101, no. 12, pp. 2498–2533, Dec 2013.
\bibitem{logic21} M. P. Kostylev, A. A. Serga, T. Schneider, B. Leven, and B. Hillebrands, “Spin-Wave Logical Gates,” Applied Physics Letters, vol. 87, no. 15, p. 153501, 2005. [Online]. Available: https://doi.org/10.1063/1.2089147
\bibitem{logic12} T. Schneider, A. A. Serga, B. Leven, B. Hillebrands, R. L. Stamps, and M. P. Kostylev, “Realization of Spin-Wave Logic Gates,” Applied Physics Letters, vol. 92, no. 2, p. 022505, 2008. [Online]. Available: https://doi.org/10.1063/1.2834714
\bibitem{logic11} K.S. Lee and S.K. Kim, “Conceptual Design of spin-wave Logic Gates Based on a Machzehnder-Type spin-wave Interferometer for Universal Logic Functions,” Journal of Applied Physics, vol. 104, no. 5, p. 053909, 2008. [Online]. Available: https://doi.org/10.1063/1.2975235
\bibitem{Magnonic_transistor} A. V. Chumak, A. A. Serga, and B. Hillebrands, "Magnon Transistor for All-Magnon Data Processing," Nature Communication, vol. 5, p. 4700, 2014.
\bibitem{logic24} B. Rana and Y. Otani, “Voltage-Controlled Reconfigurable Spin-Wave Nanochannels and Logic Devices,” Physical Review Applied, vol. 9, p. 014033, Jan 2018. [Online]. Available: https://link.aps.org/doi/10.1103/PhysRevApplied.9.014033
\bibitem{logic1} A. Khitun and K. L. Wang, “Non-Volatile Magnonic Logic Circuits Engineering,” Journal of Applied Physics, vol. 110, no. 3, p. 034306, 2011. [Online]. Available: https://doi.org/10.1063/1.3609062
\bibitem{logic19} K. Nanayakkara, A. Anferov, A. P. Jacob, S. J. Allen, and A. Kozhanov, “Cross Junction spin-wave Logic Architecture,” IEEE Transactions on Magnetics, vol. 50, no. 11, pp. 1–4, Nov 2014.
\bibitem{logic100} T. Fischer, M. Kewenig, D. A. Bozhko, A. A. Serga, I. I. Syvorotka, F. Ciubotaru, C. Adelmann, B. Hillebrands, and A. V. Chumak, "Experimental prototype of a spin-wave majority gate", Applied Physics Letter, Vol. 110, February 2017, pp.  152401-1-4.
\bibitem{logic101} F. Ciubotaru, G. Talmelli, T. Devolder, O. Zografos, M. Heyns, C. Adelmann, and I. P. Radu, "First experimental demonstration of a scalable linear majority gate based on spin waves", IEEE International Electron Devices Meeting (IEDM), January 2019, pp. 36.1.1-36.1.4.
\bibitem{logic17} I. A. Ustinova, A. A. Nikitin, A. B. Ustinov, B. A. Kalinikos, and E. Lhderanta, “Logic Gates Based on Multiferroic Microwave Interferometers,” in 2017 11th International Workshop on the Electromagnetic Compatibility of Integrated Circuits (EMCCompo), July 2017, pp. 104– 107.
\bibitem{Magnonic_crystals_for_data_processing} A. V. Chumak, A. A. Serga, and B. Hillebrands, “Magnonic Crystals for Data Processing,” Journal of Physics D: Applied Physics, vol. 50, no. 24, p. 244001, 2017. [Online]. Available: http://stacks.iop.org/0022- 3727/50/i=24/a=244001
\bibitem{LL_eq} L. Landau and E. Lifshitz., “On the Theory of the Dispersion of Magnetic Permeability in Ferromagnetic Bodies,” Physikalische Zeitschrift der Sowjetunion, pp. 101– 114, 1935.
\bibitem{G_eq} T. L. Gilbert, “A Phenomenological Theory of Damping in Ferromagnetic Materials,” IEEE Transactions on Magnetics, vol. 40, no. 6, pp. 3443– 3449, Nov 2004.
\bibitem{dispersionrelation} B. A. Kalinikos, and A. N. Slavin, "Theory of dipole-exchange spin-wave spectrum for ferromagnetic films with mixed exchange boundary conditions", Journal  Physics C: Solid State Physics, Vol. 19, 1986, pp. 7013-7033.
\bibitem{Magnetostatics_ref3} A. A. Serga, A. V. Chumak, and B. Hillebrands, “YIG Magnonics,” Journal of Physics D: Applied Physics, vol. 43, no. 26, p. 264002, 2010. [Online]. Available: http://stacks.iop.org/0022-3727/43/i=26/a=264002
\bibitem{Magnonics} V. V. Kruglyak, S. O. Demokritov, and D. Grundler, “Magnonics,” Journal of Physics D: Applied Physics, vol. 43, no. 26, p. 264001, 2010. [Online]. Available: http://stacks.iop.org/0022-3727/43/i=26/a=264001
\bibitem{parallel_data_processing1} A. Khitun, “Multi-frequency magnonic logic circuits for parallel data processing,” Journal of Applied Physics, vol. 111, no. 5, p. 054307, 2012. [Online]. Available: https://doi.org/10.1063/1.3689011
\bibitem{counter} P. SHABADI, S. N. RAJAPANDIAN, S. KHASANVIS, and C. A. MORITZ, “Design of spin-wave Functions-Based Logic Circuits,” SPIN, vol. 02, no. 03, p. 1240006, 2012. [Online]. Available: https://doi.org/10.1142/S2010324712400061
\bibitem{logic9} O. Zografos, L. Amar, P. Gaillardon, P. Raghavan, and G. D. Micheli, “Majority Logic Synthesis for spin-wave Technology,” in 2014 17th Euromicro Conference on Digital System Design, AugUST 2014, pp. 691– 694.
\bibitem{OOMMF} M. J. Donahue and D. G. Porter, “OOMMF User’s Guide, version 1.0,” Interagency Report NISTIR 6376, Sept 1999. [Online]. Available: http://math.nist.gov/oommf
\bibitem{parameters} T. Devolder, J.-V. Kim, F. Garcia-Sanchez, J. Swerts, W. Kim, S. Couet, G. Kar, and A. Furnemont, “Time-Resolved Spin-Torque Switching in MgO-Based Perpendicularly Magnetized Tunnel Junctions,” Phys. Rev. B, vol. 93, p. 024420, Jan 2016. [Online]. Available: https://link.aps.org/doi/10.1103/PhysRevB.93.024420
\bibitem{Excitation_table_ref16} O. Zografos, B. Sore, A. Vaysset, S. Cosemans, L. Amar, P. Gaillardon, G. D. Micheli, R. Lauwereins, S. Sayan, P. Raghavan, I. P. Radu, and A. Thean, “Design and Benchmarking of Hybrid CMOS-spin-wave Device Circuits Compared to 10nm CMOS,” in 2015 IEEE 15th International Conference on Nanotechnology (IEEE-NANO), July 2015, pp. 686–689.
\bibitem{Abdulqader} A. Mahmoud, F. Vanderveken, C. Adelmann, F. Ciubotaru, S. Hamdioui, and S. Cotofana, “Fan-out enabled spin wave majority gate,” AIP Advances, vol. 10, pp. 035119, March 2020. [Online]. Available: https://aip.scitation.org/doi/10.1063/1.5134690

\end{thebibliography}
\end{document}